\documentclass[usenatbib]{mnras}
\usepackage{newtxtext,newtxmath}
\usepackage[T1]{fontenc}
\usepackage{ae,aecompl}

\usepackage{graphicx}
\usepackage{amsmath}
\usepackage{amssymb}

\title[Classification of Stars in the Galactic Centre]{Random Forest Classification of Stars in the Galactic Centre}

\author[]{
P. M. Plewa$^{1}$\thanks{E-mail: pmplewa@mpe.mpg.de}
\\
$^{1}$Max-Planck-Institut f\"ur extraterrestrische Physik, Garching, Germany\\
}

\date{Accepted 2018 February 21. Received 2018 January 16; in original form 2017 September 13}
\pubyear{2018}

\begin{document}
\label{firstpage}
\pagerange{\pageref{firstpage}--\pageref{lastpage}}
\maketitle

\begin{abstract}
Near-infrared high-angular resolution imaging observations of the Milky Way's nuclear star cluster have revealed all luminous members of the existing stellar population within the central parsec. Generally, these stars are either evolved late-type giants or massive young, early-type stars. We revisit the problem of stellar classification based on intermediate-band photometry in the K-band, with the primary aim of identifying faint early-type candidate stars in the extended vicinity of the central massive black hole. A random forest classifier, trained on a subsample of spectroscopically identified stars, performs similarly well as competitive methods (${F_1=0.85}$), without involving any model of stellar spectral energy distributions. Advantages of using such a machine-trained classifier are a minimum of required calibration effort, a predictive accuracy expected to improve as more training data becomes available, and the ease of application to future, larger data sets. By applying this classifier to archive data, we are also able to reproduce the results of previous studies of the spatial distribution and the K-band luminosity function of both the early- and late-type stars.
\end{abstract}


\section{Introduction}
\label{sec:1}

The dense nuclear star cluster of the Milky Way has been observed and monitored for many years at near-infrared wavelengths, being highly extincted in the visible spectral range. Today's routine, ground-based observations at high angular resolution, assisted by adaptive optics (AO), reveal all luminous members of the existing stellar population, the composition of which has been the focus of numerous previous studies \citep[for a review, see][]{2010RvMP...82.3121G}.

Within the central parsec, the bulk of detected stars are old, evolved giants that likely formed at the same time as the Galactic bulge \citep{2011ApJ...741..108P}. However, more recently formed main-sequence stars are also detected, which are furthermore split into different sub-populations. A significant fraction of the most massive young \mbox{(WR/O-)stars} reside in a disc structure \citep{2006ApJ...643.1011P,2009ApJ...690.1463L,2009ApJ...697.1741B,2014ApJ...783..131Y}, while an apparently isotropically distributed ``S-star'' cluster of less massive \mbox{(B-)stars} is concentrated around the central massive black hole \citep{2008ApJ...689.1044G,2009ApJ...692.1075G,2016ApJ...830...17B,2017ApJ...837...30G} identified with the compact radio source Sgr~A* \citep{2007ApJ...659..378R,2015MNRAS.453.3234P}. So far, only few B-stars have been identified further out from the black hole, and it is unclear whether these belong to the stellar disk, the S-stars, or form a distinct population \citep[e.g.][]{2014ApJ...784...23M}.

To better understand the complex history of the nuclear star cluster in general, and that of the young stars in particular, it is of fundamental interest to study and characterize these different stellar populations, for example by their luminosity function, spatial distribution or kinematics. In this study, we revisit the key problem of identifying stars as members of the young or old population, by exploring alternative, machine-learning techniques for determining their spectral types from images.

The high level of stellar crowding in the Galactic Centre demands the use of integral field spectroscopy to achieve a definitive spectral classification of individual stars \citep[e.g.][]{2003ApJ...586L.127G,2005ApJ...628..246E,2009ApJ...703.1323D,2010ApJ...708..834B,2013ApJ...764..154D,2015ApJ...808..106S,2015A&A...584A...2F}. However, with respect to covering a large field of view to a sufficiently high depth in a reasonably small amount of observing time, this technique remains inefficient in comparison to imaging techniques. It is therefore of practical interest to develop accurate methods of photometric classification \citep[e.g.][]{2003ApJ...594..812G,2009A&A...499..483B,2013A&A...549A..57N}, in particular to identify the rare young, early-type candidate stars at faint magnitudes in the extended vicinity of the massive black hole. For confirmation, these stars may later be targeted sequentially in deep spectroscopic follow-up observations, of which coverage is still lacking in off-centre fields.

In section~\S\ref{sec:2} we present the intermediate-band imaging observations that allow us to determine spectral types of a few thousand stars in the nuclear cluster, before we describe the specific classification method we have used in section~\S\ref{sec:3}. In section~\S\ref{sec:4} we discuss the achieved classification performance, estimate surface density profiles of the early- and late-type stars, as well as their luminosity functions, and compare our results to those of other studies. Finally, we present our conclusions and discuss future opportunities in section~\S\ref{sec:5}.

\begin{table}
\centering
\caption{NACO/VLT Observations: For each date, we list the central wavelength~($\lambda_c$) and width~($\Delta\lambda$) of the filter used, the number of frames~(N) obtained and their exposure times~(DIT), as well as the estimated Gini importance of the respective photometry (see Sec.~\ref{sec:4.1}).}
\label{tab:1}
\begin{tabular}{lcccccc}
\hline
Date & $\lambda_c$ & $\Delta\lambda$ & N & NDIT & DIT & Importance \\
(UT) & ($\mu$m) & ($\mu$m) & & & (s) & \\
\hline
2004-07-08 & 2.00 & 0.06 & 8 & 4 & 36 & $4\%$\\
2004-06-11 & 2.06 & 0.06 & 96 & 1 & 30 & $5\%$\\
2004-06-11 & 2.24 & 0.06 & 99 & 1 & 30 & $26\%$\\
2004-07-08 & 2.27 & 0.06 & 8 & 4 & 36 & $21\%$\\
2004-07-08 & 2.30 & 0.06 & 8 & 4 & 36 & $5\%$\\
2004-06-12 & 2.33 & 0.06 & 120 & 1 & 30 & $27\%$\\
2004-07-08 & 2.36 & 0.06 & 8 & 4 & 36 & $13\%$\\
\hline
\end{tabular}
\end{table}

\section{Observations \& Data Reduction}
\label{sec:2}

The data set we make use of is a subset of the one previously analyzed by \citet{2009A&A...499..483B}, which was obtained in the year 2004 with the NACO/VLT imager and is publicly available in raw form through the ESO archive (see Tab.~\ref{tab:1}). We have re-reduced all available images using methods described in detail by \citet{2015MNRAS.453.3234P}. The image reduction process includes a standard sky-subtraction, flat-field and bad pixel corrections, as well as corrections for optical distortion and differential refraction, and a precise (sub-pixel) image registration.

\begin{figure}
\includegraphics[width=0.98\linewidth]{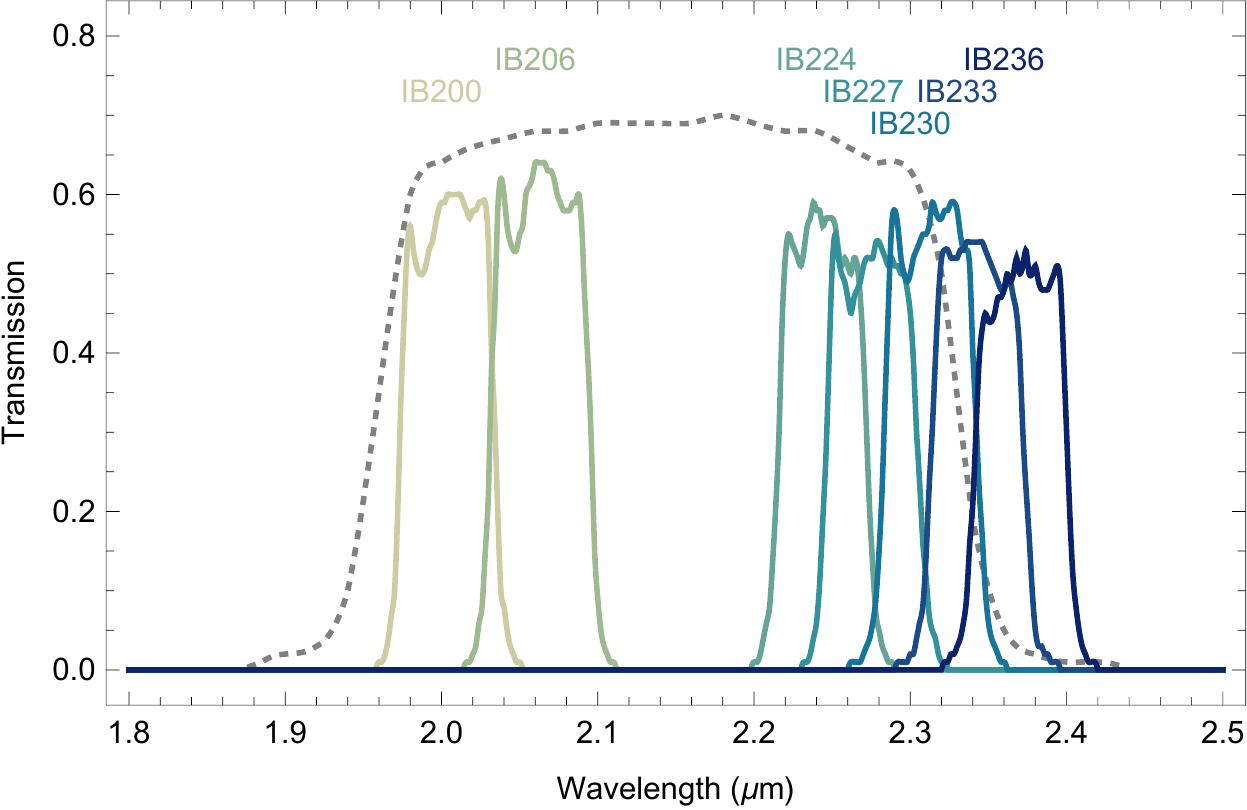}
\caption{Transmission curves of the seven intermediate-band filters used in this study, in comparison to that of the broad-band Ks filter (see \url{http://www.eso.org/sci/facilities/paranal/instruments/naco/inst/filters.html}).}
\label{fig:1}
\end{figure}

The end result of this process are seven combined images, one for each of the intermediate-band filters used, with a field of view of ${40''\times40''}$ roughly centred on Sgr~A*. The average full width at half-maximum (FWHM) of the point-spread function (PSF) is ${90\,\text{mas}}$. The two shortest filters provide a continuous coverage between wavelengths of ${1.97\mu\rm m}$ and ${2.09\mu\rm m}$, and the remaining five filters are interleaved to cover a spectral range from ${2.21\mu\rm m}$ to ${2.39\mu\rm m}$ (see Fig.~\ref{fig:1}). We have also inspected additional images taken with filters centred on wavelengths of ${2.12\mu\rm m}$ and ${2.18\mu\rm m}$ as part of the same observing program, but decided not to use them. The images taken with the former filter are of lower quality, due to poor weather conditions, whereas those taken with the latter filter are affected by recombination line emission of ionized gas in the mini-spiral streamer (Br$\gamma$). Unlike \citet{2009A&A...499..483B}, we do not use any H-band data, which is only available for a significantly more restricted field of view.

\section{Methods}
\label{sec:3}

\begin{figure*}
\includegraphics[width=0.8\linewidth]{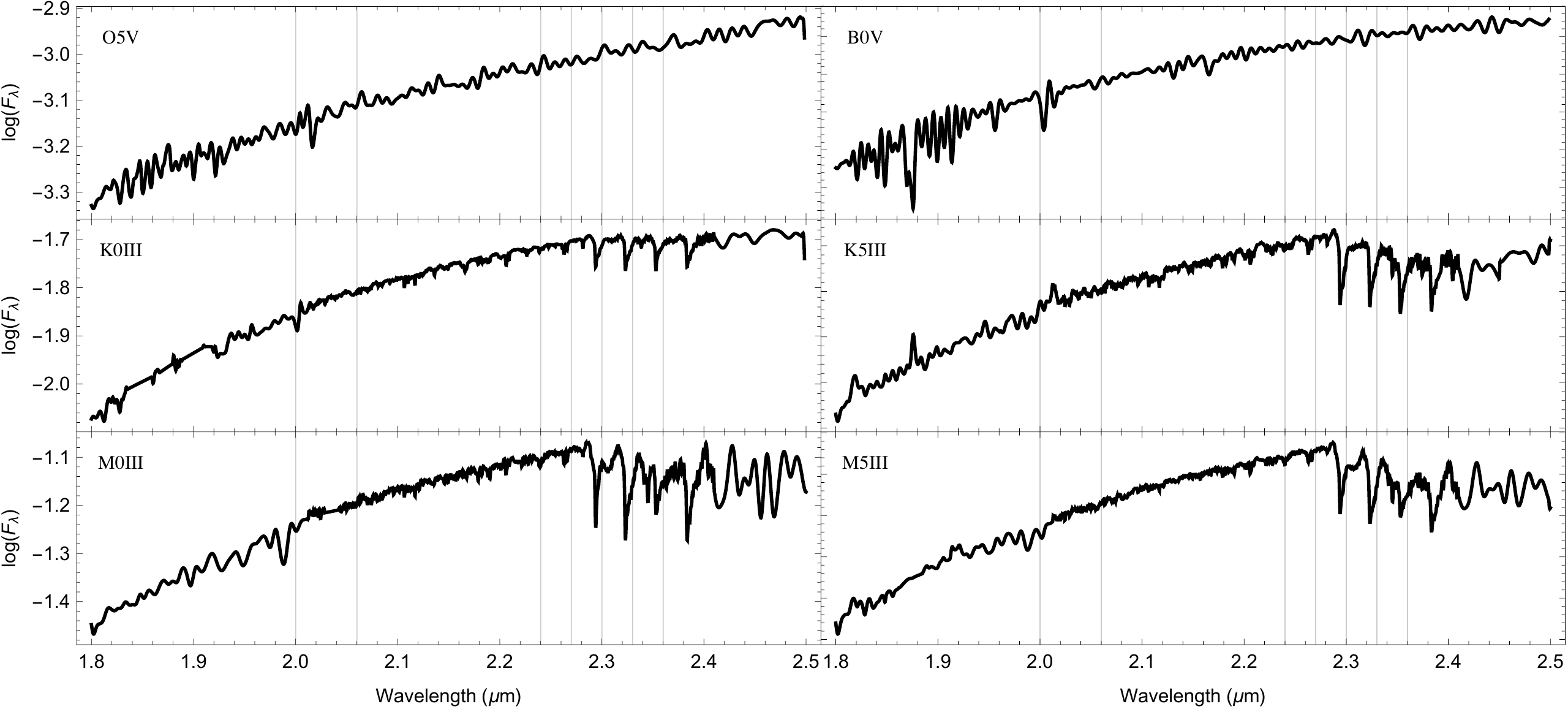}
\caption{Examples of high-resolution K-band spectra of early-type (top row) and late-type (bottom rows) stars (taken from the \citet{1998PASP..110..863P} Atlas), which are detectable in near-infrared observations of the Galactic Centre. The characteristic CO absorption features that appear in the spectra of the late-type stars allow a distinction between the two classes based on intermediate-band photometry, by sampling the stellar spectra at a few discrete points only (vertical lines, see also Fig.~\ref{fig:3}). To account for reddening, we have used the extinction law for the Galactic Centre derived by \citet{2011ApJ...737...73F}.}
\label{fig:2}
\end{figure*}

\begin{figure}
\includegraphics[width=0.98\linewidth]{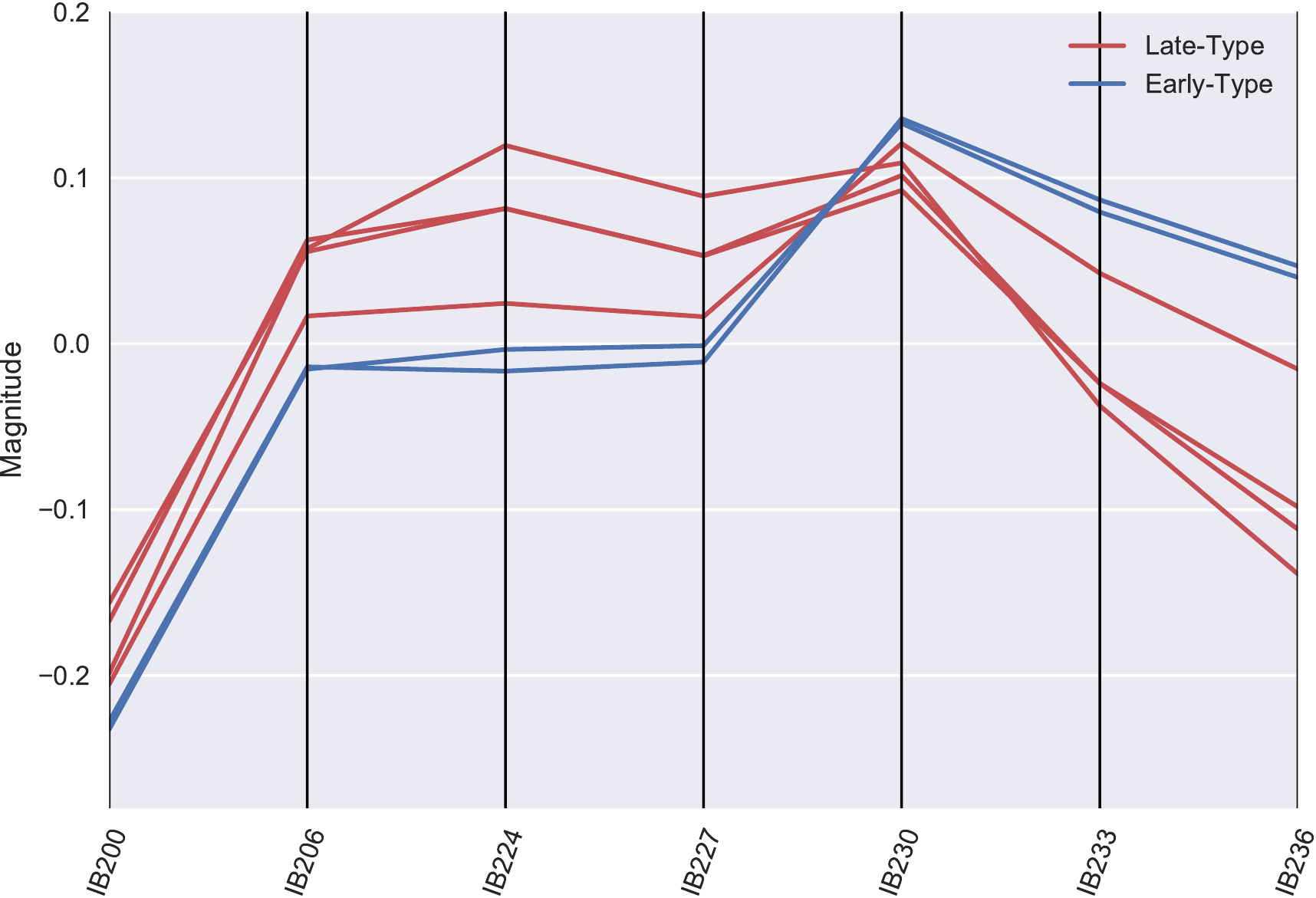}
\caption{A characteristic difference in the shape of the K-band spectrum caused by CO absorption features allows a distinction between late- and early-type stars in observations of the Galactic Centre, based on intermediate-band photometry in the K-band using the seven indicated filters, instead of a high-resolution spectrum (see also Fig.~\ref{fig:2}).}
\label{fig:3}
\end{figure}

The main spectral signature in the near-infrared K-band that facilitates a distinction between the late- and early-type stars detectable in the Galactic Centre, and allows their classification based on limited photometric information only, are distinct CO absorption features (see Fig.~\ref{fig:2}~\&~\ref{fig:3}). These features start to appear in the spectra of late G-type giants and become increasingly pronounced in giants of spectral types~K and~M. In contrast, O- and B-type main-sequence stars show an almost featureless black-body spectrum with only a few narrow, weaker absorption lines.

\begin{figure}
\includegraphics[width=0.98\linewidth]{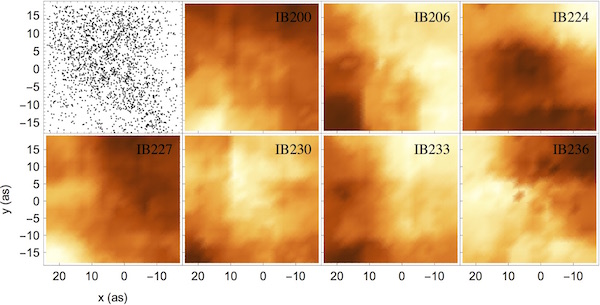}
\caption{Local calibration maps for the photometry are used to account for residual systematic features, which are reminiscent of the four-point dither pattern, as well as the spatial variability of the extinction. The first panel shows the surface density of the reference sources (see Sec.~\ref{sec:3.1}).}
\label{fig:4}
\end{figure}

\subsection{Photometry}
\label{sec:3.1}

The quality of the photometry can be expected to have a strong impact on the ultimate performance of our stellar classifier. To determine as accurately as possible the shape of each star's spectrum over the K-band, we perform PSF photometry on the seven reduced intermediate-band images using the \textit{StarFinder} tool \citep{2000A&AS..147..335D}. The \textit{StarFinder} algorithm is optimized to detect stars in a crowded field and provide accurate astrometry and photometry, when aperture photometry for instance would fail. Besides an image, the required input for the algorithm are the positions of a number of manually chosen PSF reference stars, which should ideally be relatively bright, isolated and scattered across the field of view. For each image, the final output of the algorithm is an estimate of the PSF and a star list containing the positions and instrumental magnitudes of all detected stars. We cross-match these star lists and keep only the sources detected in all seven bands, thereby removing spurious detections of faint stars. In total, we identify $3165$ sources in the field of view, with K-band magnitudes ranging from~$9.2$ to about~$16.1$ and a high completeness for sources brighter than magnitude $15.7$, at least outside the very crowded central $1''$.

The distinguishing features between the early- and late-type stars that we aim to isolate are imprinted in the spectral shape of a star and not necessarily dependent on its overall brightness. Therefore, to remove the latter information, we subtract the average value from the measured magnitudes of each star (i.e. we divide by the average flux) and in the following refer to the so-standardized multi-band photometry as a star's spectral energy distribution (SED).

A few tens of extremely bright stars in the field of view are affected by saturation in one or several bands, and a few of the prominent super-giants and WR-stars in the field are affected severely. A repair of saturated PSF cores is implemented in the \textit{StarFinder} algorithm, but the missing data may result in an increased, possibly systematic uncertainty of the SED of any saturated star. Such stars are nevertheless suitable PSF reference stars, because they provide valuable information about the extended PSF wings.

We expect another systematic uncertainty in the stellar SEDs, which is in part specific to AO-assisted observations and an inevitable consequence of the spatial and temporal variability of the PSF. The spatial variability arises due to anisoplanatism, which causes the AO correction to deteriorate at separations larger than about ${10''}$ to ${20''}$ from the AO natural guide star (GCIRS~7), as a result of differential tilt jitter \citep[e.g.][]{2010MNRAS.401.1177F}. The temporal variability arises due to changing observing conditions and performance of the AO system, from one night to another, as well as within a night. In our photometric data, the resulting effect appears similar to an intrinsic variation of the extinction across the field of view, which also exists \citep[e.g.][]{2010A&A...511A..18S,2011ApJ...737...73F}. To mitigate these effects, while continuing to use a constant PSF model, we derive an empirical, local photometric calibration following a strategy similar to that of \citet{2009A&A...499..483B}.

This local calibration relies on the fact that the early-type stars are rare in comparison to the late-type stars. We can therefore select a group of nearest neighbors around each star and use their average SED for reference at that position, such that a typical late-type star will have an approximately flat SED everywhere in the field of view. To further avoid selecting early-type stars, we only consider stars in the magnitude range from~$14.5$ to~$16$ in K-band with a minimum projected distance of $1''$ from Sgr~A* as reference sources, which are predominantly red clump stars (that produce a bump in the luminosity function at these magnitudes). Most of the excluded stars inside the central region are known to be members of the predominantly young S-star cluster \citep[e.g.][]{2017ApJ...837...30G}.

The necessary magnitude correction for each wavelength band does not affect the classification of any specific star directly, since the features of its particular SED are preserved, if only relative to the local average SED. We find that selecting a number of 20 neighboring reference stars is sufficient, which are typically distributed over a $1.7''$ circular surrounding area. The reference stars are generally not distributed evenly within that area, but any discontinuity in the calibration maps is avoided (see Fig.~\ref{fig:4}). After this calibration, the classification accuracy should not depend on a star's position in the field of view.

\subsection{Classification}
\label{sec:3.2}

We choose a machine-trained random forest classifier, which is a meta-classifier based on an ensemble of decision trees. For in-depth information about the algorithm and details about the implementation, we refer the reader to \citet{Breiman:2001fb} and \citet[][see also \url{http://scikit-learn.org/}]{Pedregosa:2011tv}.

A decision tree has several advantages as a method of classification. It can essentially be reduced to a set of rules that, in our case, can be used to predict the class (i.e. spectral type) of a star from its SED, or to obtain class membership probabilities, which are straightforward to interpret. Generally, there is also little data preparation required, for example neither rescaling nor feature selection, which is performed implicitly. In our case, neither an absolute nor even a relative magnitude calibration for the individual wavelength bands is strictly necessary, due to the mentioned scaling invariance, and because the training and test sets used for fitting and evaluating the classifier are subsets of the same data set. Furthermore, the computational cost of performing a classification using a decision tree does not depend strongly on the size of the training set (but scales logarithmically). The main disadvantage of a decision tree is a susceptibility to over-fitting and instability with respect to changes in the training set. This is mitigated by constructing an ensemble of decision trees, for instance a random forest, where multiple trees are fit to random subsamples of the data and results are averaged, to improve the robustness and the overall predictive accuracy of the classification.

It is important to ensure that the stars included in the training set have representative SEDs for each class and that their classification is indeed correct. We join a sample of early-type stars reported by \citet{2014ApJ...783..131Y}, which includes data originally published by \citet{2006ApJ...643.1011P,2009ApJ...697.1741B,2009ApJ...703.1323D} and \citet{2013ApJ...764..154D}, and a sample of late-type stars reported by \citet{2007ApJ...669.1024M}. These stars were manually classified using high-resolution near-infrared spectroscopic data (obtained with the SINFONI/VLT and OSIRIS/Keck spectrographs), based on the presence of CO absorption features or narrow absorption lines of HI (Br$\gamma$), HeI or NaI in the stellar spectra. However, we find that the two samples have two sources in common, for which we trust the more recent classification as early-type stars. The late-type sample is widely distributed within the field of view, yet all observed fields are located north of Sgr~A*. The early-type sample is concentrated around Sgr~A*, being limited in size by the coverage of deep spectroscopic observations. In total, we were able to match $114$ of the $116$ early-type stars and $215$ of the late-type stars to our star list, that comprise our training set. One of the missing early-type stars is extremely saturated, the other is extremely faint.

\section{Results}
\label{sec:4}

\begin{figure}
\includegraphics[width=0.98\linewidth]{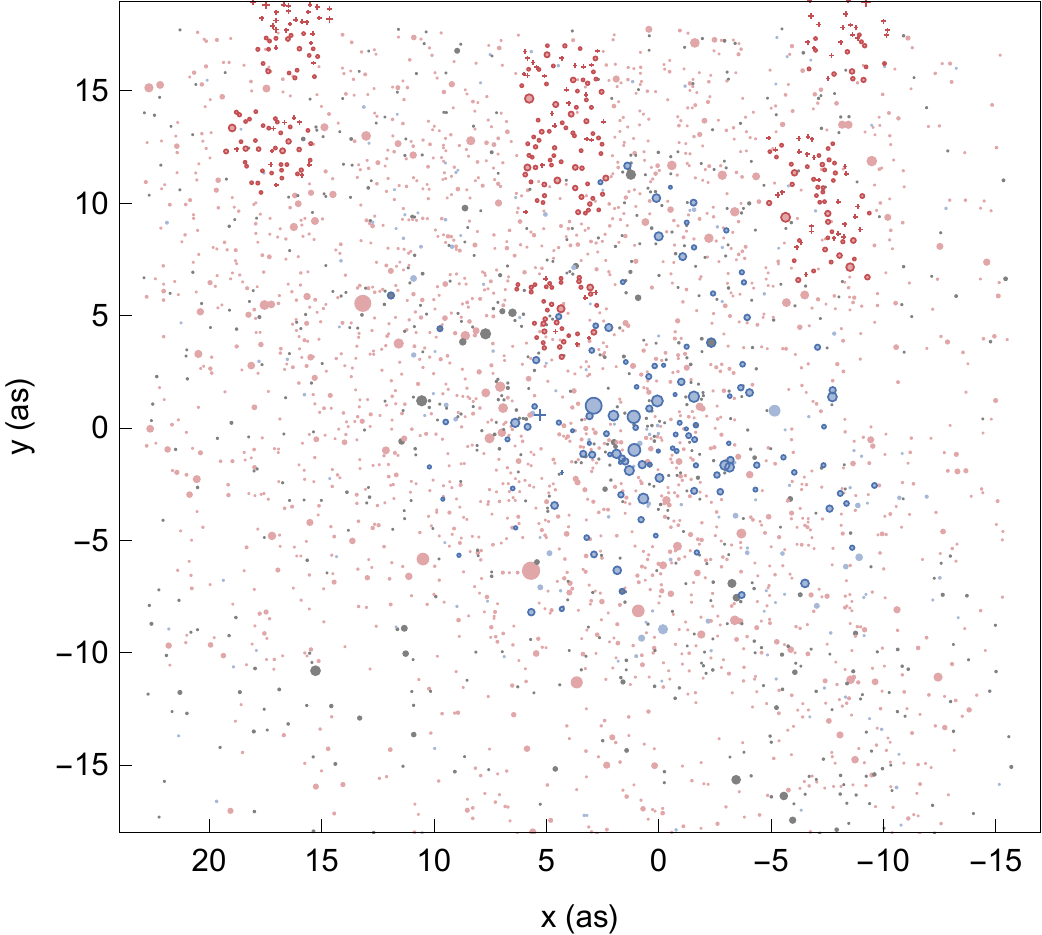}
\caption{A visualization of the classification results, where early- and late-type candidate stars are shown in blue and red colour, and unclassified stars are shown in grey, at their angular separations from Sgr~A*. The stars that comprise the training set are highlighted in a darker colour.}
\label{fig:5}
\end{figure}

\begin{figure}
\includegraphics[width=0.98\linewidth]{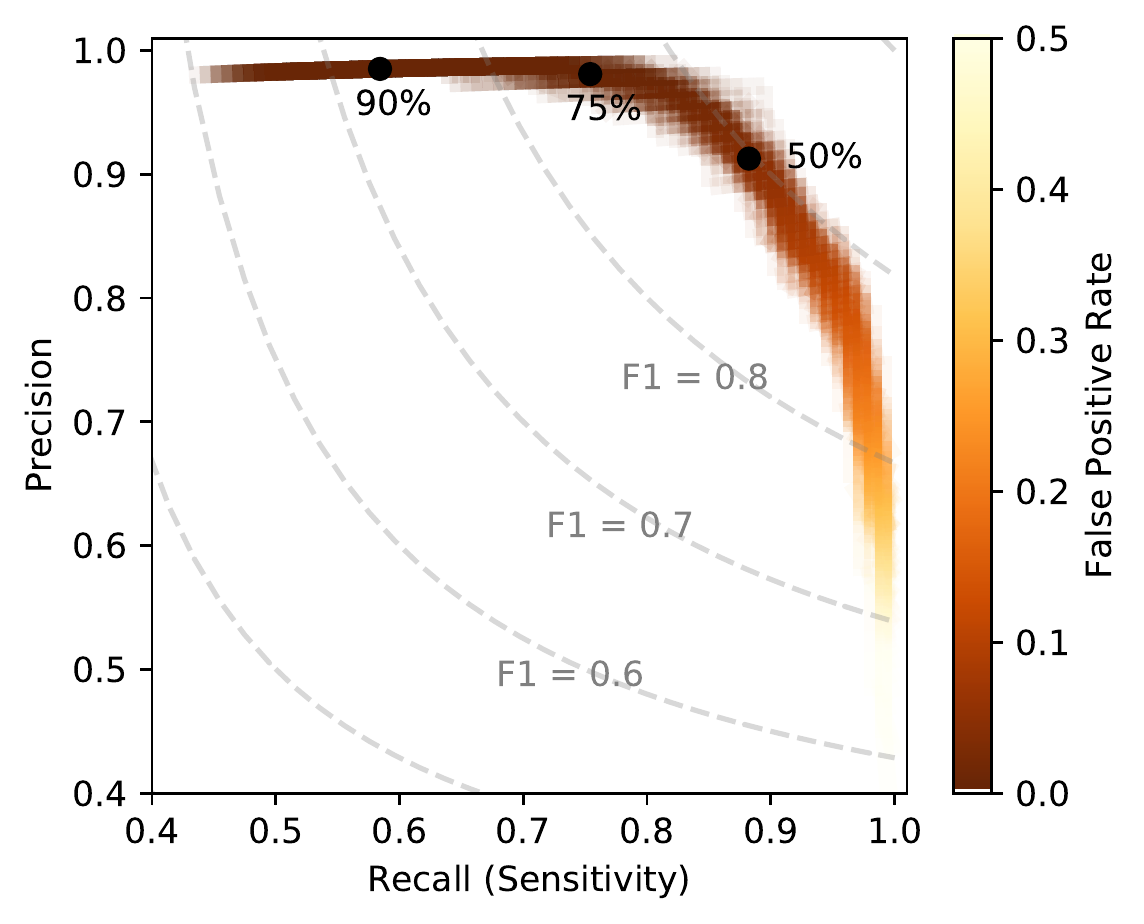}
\caption{Performance of the random-forest classifier with respect to identifying early-type candidate stars, estimated using cross-validation. The labels of the highlighted points indicate the respective thresholds for the class membership probability (see text).}
\label{fig:6}
\end{figure}

\begin{table*}
\caption{Classification Results (abridged). Type: Type of the star, if it is part of the training set, i.e. if it has been classified spectroscopically (E: early-type star, L: late-type star). x/y: Angular separation of the star from Sgr~A* in standard coordinates, where x and y increase in the directions of East and North. IB200 - IB236: Instrumental (\textit{StarFinder}) magnitudes in the respective bands (see also Fig.~\ref{fig:1}). The K-band magnitude of each star can be calculated approximately as ${\text{K}\approx\text{IB224}+24.63}$ \citep[see][for the photometric standards used]{2007ApJ...659.1241R}. P(E): Estimated probability for the star being an early-type star, where ${P(L)=1-P(E)}$. For stars in the training set, cross-validated estimates are given. The table is sorted by apparent brightness in the IB224 filter.}
\label{tab:2}
\begin{tabular}{llrrrrrrrrrr}
\hline
Type & x (as) & y (as) & IB200 & IB206 & IB224 & IB227 & IB230 & IB233 & IB236 & P(E) \\
\hline
     &  5.676 &  -6.351 & -13.84 & -14.39 & -15.39 & -15.71 & -15.56 & -14.51 & -14.74 & 0.13 \\
     & 13.161 &   5.543 & -13.89 & -14.42 & -15.21 & -15.43 & -15.27 & -14.79 & -14.59 & 0.06 \\
   E &  2.892 &   0.989 & -14.26 & -15.03 & -15.07 & -15.23 & -15.27 & -14.89 & -15.05 & 0.89 \\
     &  0.908 &  -8.138 & -13.13 & -14.08 & -14.46 & -14.70 & -14.49 & -14.02 & -14.05 & 0.01 \\
     & 10.486 &  -5.832 & -12.74 & -13.71 & -14.46 & -14.62 & -14.35 & -14.12 & -13.99 & 0.20 \\
   E &  1.108 &   0.496 & -13.49 & -14.39 & -14.40 & -14.38 & -14.38 & -14.18 & -14.22 & 0.76 \\
     &  3.645 & -11.318 & -12.84 & -13.87 & -14.33 & -14.26 & -14.36 & -13.82 & -13.61 & 0.05 \\
   E &  1.083 &  -0.978 & -13.29 & -14.20 & -14.31 & -13.92 & -14.12 & -14.15 & -14.02 & 0.76 \\
     & -5.160 &   0.767 & -13.00 & -13.75 & -14.26 & -14.50 & -14.58 & -14.27 & -14.44 & 0.95 \\
     & 10.541 &   1.209 & -13.05 & -13.92 & -14.15 & -14.18 & -14.12 & -13.93 & -13.77 & 0.34 \\
\hline
\end{tabular}
\end{table*}

\subsection{Classification Performance}
\label{sec:4.1}

Due to the small total number of spectrally classified stars, it is infeasible to keep back a dedicated, representative validation set without compromising the ability to train our classifier. We instead evaluate the classifier's performance by conducting ($10$-fold) stratified cross-validation on the training set. This means splitting the training set into complementary subsets, accounting for class imbalance, and repeatedly training and testing the classifier using one of the subsets for validation, while using the others combined for training. Each time, the (hyper-)parameters of the classifier are re-optimized as well, using a random search to minimize cross-entropy loss, which is estimated by conducting a second round of cross-validation on the provisional training sets. This whole nested cross-validation procedure is repeated multiple times, to obtain unbiased estimates of the classifier's average performance metrics and their uncertainty \citep[e.g.][]{Cawley:2010ue}. The main parameters to be optimized are the number of features to consider when splitting a tree (${N_\text{features}\approx\sqrt{N_\text{filters}}}$), the depth of the trees (${N_\text{samples, min.}\gtrapprox1}$ at each leaf), and the number of trees in the random forest ensemble (${N_\text{trees}\approx300}$). For completeness, we have also checked that the out-of-bag error rates for selected parameter combinations are reasonable (approx. 7\%), which can be estimated already at the training stage.

In total, $274$ sources in the field of view are classified as early-type stars (class $E$) and $2216$ as late-type stars (class $L$), each with an estimated class membership probability of at least $75\%$ (see Fig.~\ref{fig:5} \& Tab.~\ref{tab:2}). Of the former, $60$ are B-type candidate stars to be confirmed spectroscopically (with ${14.5\lesssim K\lesssim15.5}$). The remaining $675$ candidate sources could not be classified reliably using this probability threshold (i.e. have class membership probabilities ${P(E)<75\%}$ and ${P(L)=1-P(E)<75\%}$). The classification of late-type stars is more reliable overall, since roughly $80\%$ of them have a class membership probability exceeding $90\%$, compared to $60\%$ of the early-type stars.

Based on the cross-validated confusion matrices, we find that the classifier has a high expected overall accuracy of $84\%$. With respect to identifying early-type stars, the sensitivity (or recall) is $75\%$ and the precision is $98\%$ (${F_1=0.85}$). Regarding late-type stars, the respective numbers are $89\%$ and $97\%$ (${F_1=0.93}$). Again, we have required a minimum class membership probability of $75\%$. This probability threshold could be adjusted downwards to trade precision for better sensitivity, but at the cost of increasing the false positive detection rate for early-type stars to above the $1\%$ level, which we specifically try to avoid (see Fig.~\ref{fig:6}). When using a relatively high threshold value, the early- or late-type stars that are not identified as such are only rarely assigned the wrong spectral type, but instead remain unclassified.

To enable a comparison with the study by \citet{2009A&A...499..483B}, we have also cross-matched their star list with our training set. We find that their method of classifying of early-type stars, which involves a direct modeling of the stellar SEDs, appears to be somewhat more reliable, judging on an achieved sensitivity of $85\%$ and a precision of $100\%$ (${F_1=0.92}$), when considering this common subset of stars. However, the full star lists, when cross-matched, differ in as many as $28\%$ of cases ($739$ sources). The majority of these different predictions involve stars that are not confidently classified as early- or late type stars in either list, but the star list of \citet{2009A&A...499..483B} notably contains $59$ stars labeled late-type that we have classified as early-type stars, and $18$ stars labeled early-type that we have classified as late-type.

Of the seven intermediate-band filters, the ones centred on wavelengths of ${2.33\mu\rm m}$, ${2.24\mu\rm m}$ and ${2.27\mu\rm m}$ prove to be the most critical for the purpose of classifying early- and late-type stars in terms of the Gini importance (see Tab.~\ref{tab:1}), which can be estimated as part of the classifier's training process. The ${2.36\mu\rm m}$ filter provides valuable information as well, but the other filters are less essential. This empirical ranking matches our expectations regarding CO absorption being the distinguishing feature (see Fig.~\ref{fig:3}), but also accounts for variance in the photometric quality across filters.

\subsection{The Stellar Population}
\label{sec:4.2}

\begin{figure}
\includegraphics[width=\linewidth]{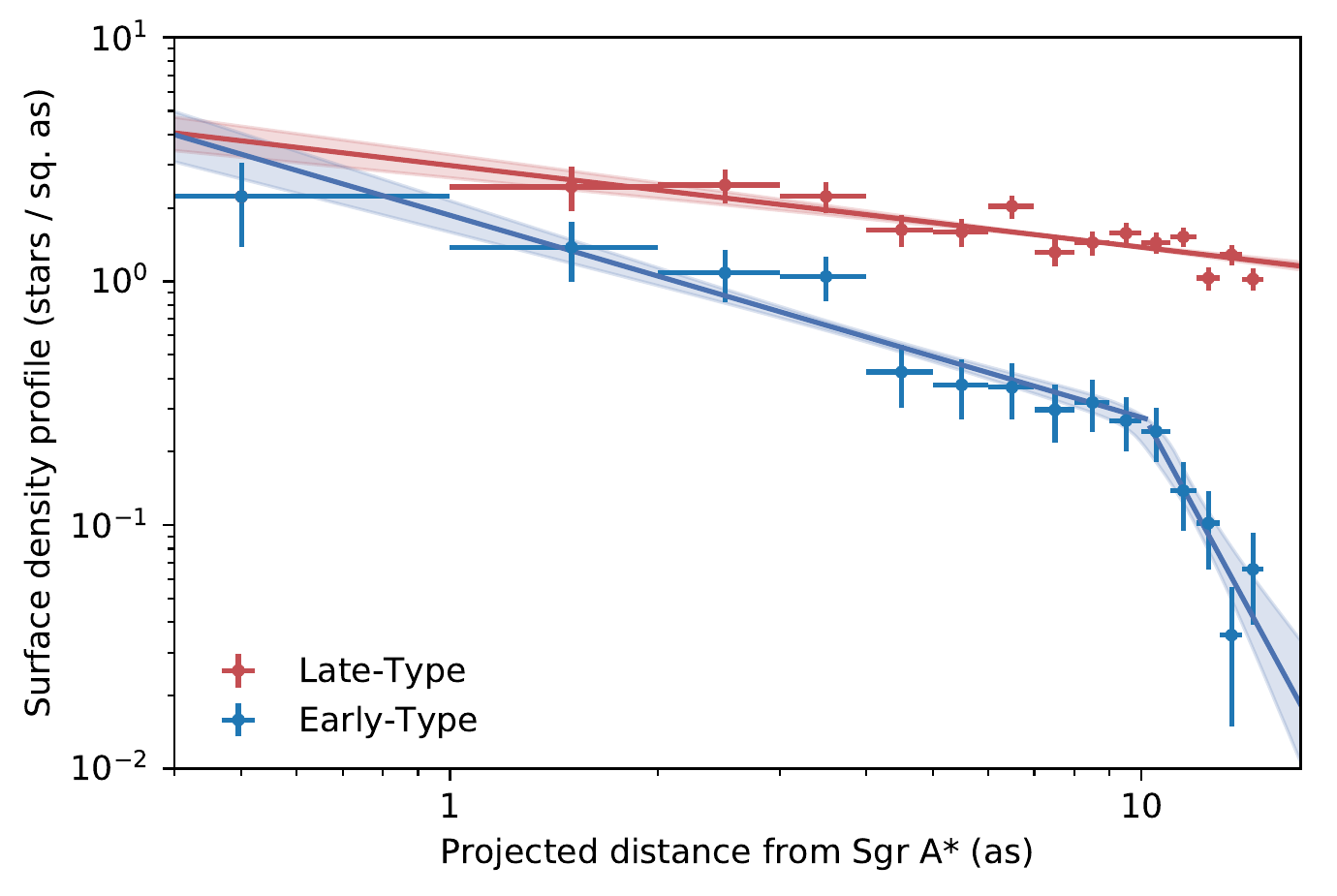}
\includegraphics[width=\linewidth]{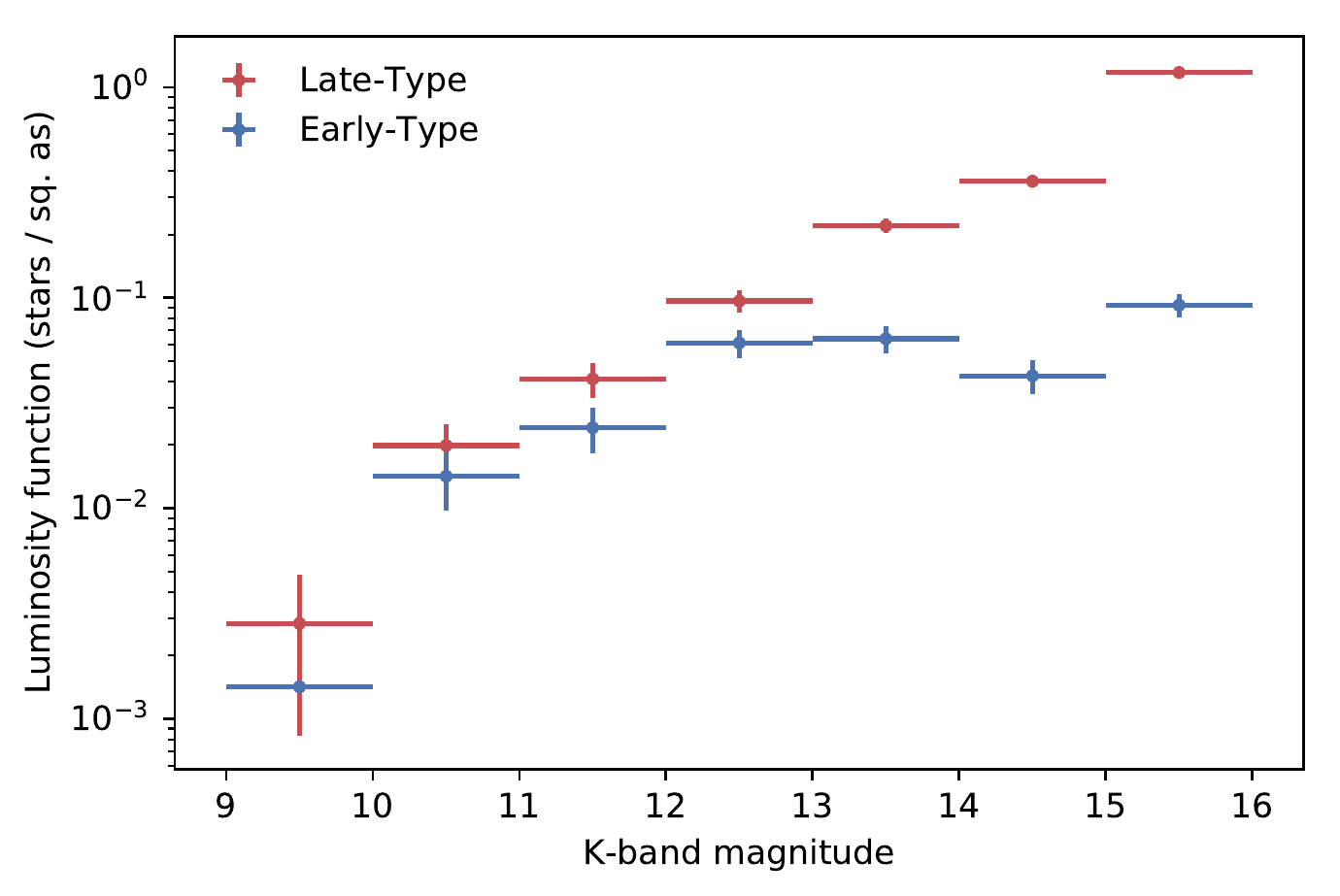}
\caption{The surface density profile (top panel) and the K-band luminosity function (bottom panel) of early- and late-type stars in the Galactic Centre. The points mark the stellar number counts in distance and magnitude bins, as indicated by the horizontal error bars, and the vertical error bars indicate Poisson standard errors of the bin counts. For model fitting, the unbinned values are used (see Sec.~\ref{sec:4.2}).}
\label{fig:7}
\end{figure}

Apart from the identification of candidate early-type stars for follow-up spectroscopic observations, or promising fields, the large-scale spatial distribution and the luminosity function of the early- and late-type stars are of immediate interest, which we are able to re-estimate using our stellar classification (see Fig.~\ref{fig:7}).

We can reproduce and confirm the results of \citet{2009A&A...499..483B} and other studies of the spatial distribution of stars in the Galactic Centre \citep[e.g.][]{2006ApJ...643.1011P,2009ApJ...690.1463L,2009ApJ...697.1741B,2009ApJ...703.1323D,2013ApJ...764..154D}. Following \citet[][Appendix~E]{2013ApJ...764..154D}, we determine the respective surface density profiles by means of Bayesian inference, using a power-law profile as a model (${\Sigma\propto R^\alpha}$). This approach does not require binning the star counts, and allows taking into account the estimated class membership probabilities as weights. For the late-type stars, we find a rather flat density profile (${\alpha=-0.33\pm0.05}$), which appears essentially core-like towards the very centre. This still presents a puzzle, because these stars are old enough to be relaxed and would be expected to form a significantly steeper \citet{1976ApJ...209..214B} cusp in that case \citep[${-0.75<\alpha<-0.5}$, but see][]{2017arXiv170103816G,2017arXiv170103817S}. For the early-type stars, we find a broken power-law profile with a turnover radius at about ${10''}$, which we interpret as the outer edge of the young stellar disc \citep[see][]{2015ApJ...808..106S}. Within that radius, the surface density profile has a well-defined index of ${\alpha=-0.83\pm0.10}$, or ${\alpha=-1.00\pm0.15}$ if we exclude the central S-star cluster (at ${R<1''}$). Beyond that radius, the number of early-type stars drops rapidly.

We can also reproduce the results of previous studies focused on the stellar K-band luminosity function (KLF). For the the late-type stars with ${K<14.5}$, i.e. if we exclude red clump stars, we find a KLF well-described by a relation $N\propto10^{\beta K}$ with an index of ${\beta=0.36\pm0.02}$, which is similar to that of the Galactic bulge \citep[e.g.][]{2003ApJ...594..812G,2009A&A...499..483B}.
The KLF of the early-type stars appears to be top-heavy in the disc region (${1''\lesssim R\lesssim10''}$), as also found by \citet{2006ApJ...643.1011P}, \citet{2009A&A...499..483B} and \citet{2010ApJ...708..834B}. The question of whether the KLF of these stars is indeed flat or somewhat steeper at the faint end, as found by \citet{2013ApJ...764..154D}, can likely be answered conclusively only on the basis of spectroscopic observations with a high completeness for B-stars over the entire region, which is currently only reached by deep imaging observations (e.g. $\gtrsim90\%$).

\section{Conclusions}
\label{sec:5}

For a proof of concept, we have constructed a machine-trained, random forest classifier to identify early- and late-type stars in the environment of the Galactic Centre black hole, based on intermediate-band photometry using seven filters in the near-infrared K-band.

\begin{figure}
\includegraphics[width=0.98\linewidth]{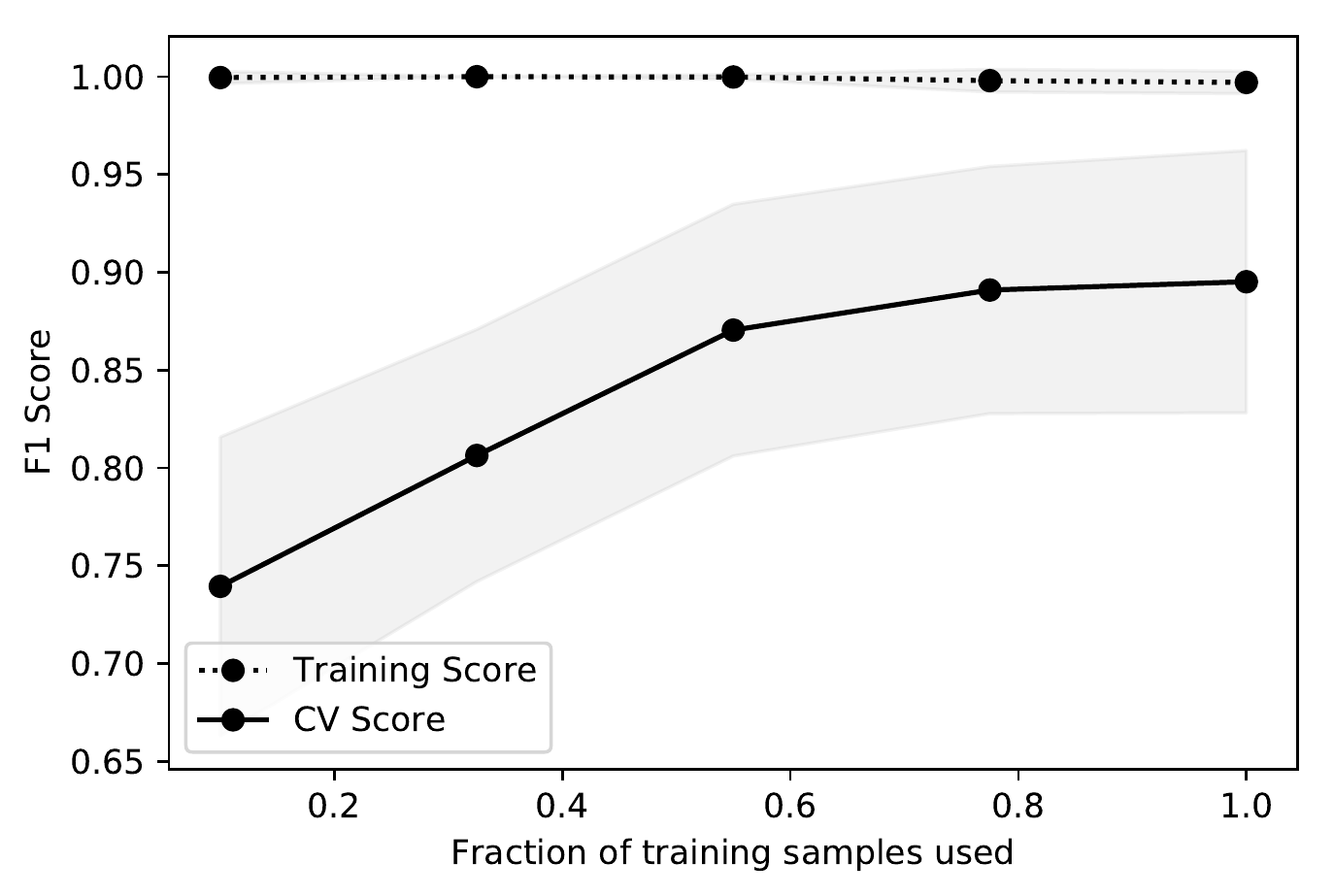}
\caption{Learning curve of the random forest classifier. The performance as measured by the cross-validated F1 score (with respect to identifying early-type stars, and assuming a class membership probability threshold of 50\%; see also Fig.~\ref{fig:6}) could probably be increased still by using a larger or higher-quality training set.}
\label{fig:8}
\end{figure}

With respect to identifying early-type candidate stars in particular, we have demonstrated that our classifier performs similarly well as competitive methods (${F_1=0.85}$), and we have identified $60$ favorable stars in the field of view for follow-up study. The classifier requires an existing training set of spectroscopically classified stars, but requires neither a model of stellar SEDs, nor a calibration of stellar magnitudes beyond accurate instrumental photometry. In principle, a machine-trained classifier will also exploit even subtle class-distinguishing features in the SEDs, i.e. spectral features other than CO absorption, which could make it highly accurate, although such features would not be interpretable in the context of some physically motivated SED model. However, to improve on methods based on direct modeling \citep[e.g.][]{2009A&A...499..483B}, a larger or higher-quality training set would probably be necessary (see Fig.~\ref{fig:8}). Increasing the spectral coverage further, by using additional filters, could be beneficial as well, as would be improvements in the photometric precision and accuracy.

In particular, we find that the brightness of the early- and late-type stars differs in the important filters by only $0.05$ to $0.1$ magnitudes (see also Fig.~\ref{fig:3}). Even though it is the combined multi-band photometry that makes a spectral classification feasible, to achieve the best possible accuracy, the photometric precision must be sufficiently high in comparison (considering also the intrinsic photometric scatter due to the diversity of spectral types), and a similarly high photometric accuracy must be guaranteed over the whole field of view. While the necessary levels of precision and accuracy have been demonstrated repeatedly \citep[e.g.][]{2010SPIE.7736E..1IL,2010A&A...509A..58S}, capabilities for high-precision photometry, specifically in crowded fields, will be greatly improved when the first instruments on the next generation of large optical telescopes will commence routine operations. We expect that the MICADO imager \citep{2016SPIE.9908E..1ZD}, for example, exploiting the unique light-collecting and resolving power of the ELT, will provide excellent photometry for a much larger sample of stars in the nuclear cluster than is presently possible to obtain. The currently limiting uncertainties in determining the PSF will be reduced by employing an advanced multi-conjugate AO system \citep[MAORY, see][]{2016SPIE.9909E..2DD}, as well as developing complementary PSF reconstruction and modeling techniques as part of the instrument design process. Also, despite not having an angular resolution as high, JWST will likely provide high-quality photometry of the nuclear star cluster as well, due to the PSF stability resulting from the stable conditions of its space environment.

As monitoring observations of the Galactic Centre continue, several more early-type stars are likely to be identified spectroscopically over the next few years already, and any further improvements in spectroscopic sensitivity will also help to grow and clean photometric training sets for stellar classification. We are therefore confident that a machine-trained classifier will be useful when applied to future, larger data sets of the Galactic Centre, i.e. deep wide-field imaging observations, even if it is only for an efficient preliminary stellar classification.

\section*{Acknowledgements}

We thank the anonymous reviewer for valuable and helpful comments and suggestions.



\bibliographystyle{mnras}
\bibliography{references}


\bsp 
\label{lastpage}
\end{document}